\documentclass[11pt,superscriptaddress,aps,prd,preprint]{revtex4}
\usepackage{amsmath}

\makeatletter
\usepackage[T1]{fontenc}
\usepackage{amsmath}
\usepackage{amssymb}
\usepackage{graphicx}

\usepackage{color}

\usepackage{slashed}

\DeclareSymbolFont{extraitalic}      {U}{zavm}{m}{it}
\DeclareMathSymbol{\Qoppa}{\mathord}{extraitalic}{161}
\DeclareMathSymbol{\qoppa}{\mathord}{extraitalic}{162}
\DeclareMathSymbol{\Stigma}{\mathord}{extraitalic}{167}
\DeclareMathSymbol{\Sampi}{\mathord}{extraitalic}{165}
\DeclareMathSymbol{\sampi}{\mathord}{extraitalic}{166}
\DeclareMathSymbol{\stigma}{\mathord}{extraitalic}{168}

\newcommand{\bea}{\begin{eqnarray}}
\newcommand{\eea}{\end{eqnarray}}

\newcommand{\pr}[1]{\ensuremath{\left[#1\right]}}
\newcommand{\pc}[1]{\ensuremath{\left(#1\right)}}
\newcommand{\chav}[1]{\ensuremath{\left\{#1\right\}}}

\begin{document}

\title{G\"{o}del and G\"{o}del-type solutions in the Palatini $f(R,T)$ gravity theory  }

\author{J. S. Gon\c{c}alves}\email[]{junior@fisica.ufmt.br }
\affiliation{Instituto de F\'{\i}sica, Universidade Federal de Mato Grosso,\\
78060-900, Cuiab\'{a}, Mato Grosso, Brazil}

\author{A. F. Santos}\email[]{alesandroferreira@fisica.ufmt.br}
\affiliation{Instituto de F\'{\i}sica, Universidade Federal de Mato Grosso,\\
78060-900, Cuiab\'{a}, Mato Grosso, Brazil}

\begin{abstract}

The Palatini $f(R,T)$ gravity theory is considered. The standard Einstein-Hilbert action is replaced by an arbitrary function of the Ricci scalar $R$ and of the trace $T$ of the energy-momentum tensor. In the Palatini approach, the Ricci scalar is a function of the metric and the connection. These two quantities, metric and connection, are taken as independents variables. Then, it is examined whether Palatini $f(R,T)$ gravity theory allows solutions in which lead to violation of causality. The G\"{o}del and G\"{o}del-type space-times are considered. In addition, a critical radius, which permits to examine limits for violation of causality, is calculated.  It is shown that, for different matter contents, non-causal solutions can be avoided in this Palatini gravitational theory.

\end{abstract}

\maketitle

\section{Introduction}

Recent observational data suggest that the expansion of the universe is accelerating. These data are confirmed by different sources, such as cosmic observations from Supernovae Ia \cite{expansao_ace, expansao_ace2}, Cosmic Microwave Background (CMB) radiation \cite{WMAP2007, WMAP2009, WMAP2011}, Large Scale Structure (LSS) \cite{LSS_evidence}, weak lensing \cite{weak_lensing}, among others. These observations are not properly explained by General Relatvity (GR). A natural question then arises: how can this acceleration be explained? To reconcile observations data and a theory that describes gravity, two main approaches are considered. That is, the observed accelerated expansion of the universe is due to some kind of extra fluid-like contribution, known as dark energy \cite{dark_energy1,dark_energy2,k_essece_1}, or GR itself is modified \cite{review1}. These modified theories represent a generalization of GR where some combination of curvature invariants (such as the Riemann tensor, the Weyl tensor, the Ricci tensor, among others) replaces or is added into the classical Hilbert-Einstein action, that is composed by the Ricci scalar term. For a review of modified theories of gravity, see \cite{review1, review2, review3, review4}. In this paper, the $f(R,T)$ gravity theory is explored.

There are several attempts to modify GR. One possibility is the $f(R,L_m)$ gravity theory which assumes that the geometric and the matter terms in the Einstein-Hilbert action are modified. This gravitational theory has been the first model to consider this coupling. In essence, it consists of replacing the Einstein-Hilbert action with an arbitrary function of the Ricci scalar $R$ and the matter Lagrangian $L_m$ \cite{RL, RL1}. Another generalization is the $f(R,T)$ gravity \cite{Harko}. In this model, the field equations are obtained from an arbitrary function that depends on the Ricci scalar $R$ and $T$, the trace of the energy-momentum tensor $T_{\mu\nu}$. The addition of $T$ in the gravitational Lagrangian leads to possible investigations of a quantum description of gravity from the $f(R,T)$ theory \cite{QT}. In recent years, this modified gravitational theory has been intensively studied \cite{fRT1, fRT2, fRT3, Houndjo, Momeni, Shabani, Sharif, Kiani, Sharif2, Jamil, Sharif3, Azizi, Alvaro, Reddy, Moraes}.

The gravitational field equations can be obtained by two different approaches, namely the metric and the Palatini formalisms.  The metric approach takes the Levi-Civita connection as the connection, and the action is varied with respect to the metric. The Palatini formalism has been introduced by A. Einstein \cite{Einstein1, Einstein2}. In this approach, the metric and connection are treated as independents fields. In GR both formalisms lead to the same field equations. However, in modified gravity theories, different field equations are obtained. It is important to note that, the order of the field equations is different. In the metric formalism, the field equations are higher-order derivatives, while in the Palatini approach the field equations are second-order derivatives. For a review of Palatini formalism applied to modified gravity, see \cite{palatini_motivation1}. In this paper, the main objective is to explore the causality question in the $f(R,T)$ gravity theory formulated in Palatini formalism \cite{fRT_Palatini_primeiro, palatini_fRT1}. 

Causality and chronology are fundamental elements in the theory of special relativity. In this theory, the chronology is preserved and causality is respected. From a local point of view, GR has the same causal structure as special relativity, since GR space-times are locally Minkowskian. However, on a global scale, the field equations of GR do not require non-local constraints on the space-times, then interesting differences can arise. There are solutions to the GR field equations that present violation of causality in the form of Closed Timelike Curves (CTCs). As an example, the G\"{o}del solution can be considered. This cosmological model shows that, although GR is local Lorentzian which leads to the local validity of the causality principle, it admits solutions with CTCs. In GR it is known that there are other solutions to the field equations that lead to the violation of causality \cite{godel,ctc1, ctc2}. Here, the gravity is governed by the $f(R,T)$ gravity theory, constructed in the Palatini approach. Then various issues must be reexamined in its framework, including the question of whether this gravity theory permits the violation of causality, which is permitted in general relativity. In this paper, the G\"{o}del solution is considered. This cosmological solution of GR has been proposed by K. G\"{o}del \cite{godel}. It is an exact solution of GR with cosmological constant $\Lambda$ which leads to the possibility of CTCs, which allow violation of causality.

Some years later of the original solution, a G\"{o}del-type metric has been developed \cite{tipo_godel1}.  This metric provides more information about the existence of CTCs. From this solution, a critical radius $r_c$, beyond which the causality is violated, can be constructed. The causality problem has been analyzed in various modified theories of gravity \cite{fr_and_godel, kessence_and_godel, chersimon_and_godel1, chersimon_and_godel2, ft_and_godel, frt_and_godel, bumblebeee_and_godel, horava_and_godel, brans_and_godel, frq_and_godel, godel_fRT, tipo_godel_fR, palatini_fR}. Considering different contents of matter, such as the perfect fluid and the scalar field, the question of breakdown of causality in Palatini $f(R, T)$ gravity theory is examined. 

The present paper is organized as follows. In section II, the field equations of Palatini $f(R,T)$ gravity are derived. In section III, the G\"{o}del solution is considered. The field equations show that the violation of causality is permitted in this gravitational model. In section IV, the G\"{o}del-type solution is introduced. The gravitational equations are solved for three different matter contents: (i) perfect fluid; (ii) perfect fluid plus scalar field; (iii) only scalar field. In addition, the critical radius is analyzed for different situations. In section V, some remarks and conclusions are presented.

\section{Palatini formulation of $f(R,T)$ gravity}

In this section, field equations for Palatini $f(R,T)$ gravity theory are obtained. In the Palatini formalism, the curvature scalar is regarded as a function of the metric tensor and the connection, i.e. $R(g,\tilde{\Gamma})$. Then the gravitational action of Palatini $f(R,T)$ gravity is given as
\begin{equation}\label{action_1}
    S=\frac{1}{2\kappa^2}\int d^4x \sqrt{-g}\, f\left(R(g,\tilde{\Gamma}), T\right) + \int d^4x \sqrt{-g}\, \mathcal{L}_m(g,\psi),
\end{equation}
where $\kappa^2 = 8\pi$, $g=det(g_{\mu\nu})$ and $f$ is a function of the Ricci scalar and the trace of the energy-momentum tensor $T_{\mu\nu}$. The matter Lagrangian $\mathcal{L}_m(g,\psi)$ is a function of $g$ and of the physical fields $\psi$. The Ricci scalar dependent on $g$ and of the Palatini connection $\tilde{\Gamma}$ is written as
\begin{equation}
    R\left(g,\tilde{\Gamma}\right) =g^{\mu\nu} \tilde{R}_{\mu\nu}\left(\tilde{\Gamma}\right),  
\end{equation}
with $\tilde{R}_{\mu\nu}\left(\tilde{\Gamma}\right)$ being the Ricci tensor expressed only in terms of the Palatini connection. It is defined as
\begin{equation}
\tilde{R}_{\mu\nu} = \partial_{\lambda} \tilde{\Gamma}_{\mu\nu}^\lambda - \partial_{\nu} \tilde{\Gamma}_{\mu\lambda}^\lambda + \tilde{\Gamma}_{\mu\nu}^\lambda \tilde{\Gamma}_{\lambda \alpha}^\alpha - \tilde{\Gamma}_{\mu\lambda}^\alpha \tilde{\Gamma}_{\nu \alpha}^\lambda,
\end{equation}
where $\tilde{\Gamma}$ is a quantity to be determined.

By taking the matter Lagrangian to be independent of $\partial_\lambda g_{\mu\nu}$, the energy momentum tensor is obtained as
\begin{equation}\label{tensor energia}
    T_{\mu\nu} = \frac{-2}{\sqrt{-g}} \frac{\partial \pc{\sqrt{-g}\mathcal{L}_m}}{\partial g^{\mu\nu}} = -2 \frac{\partial \mathcal{L}_m}{\partial g^{\mu\nu}} + g_{\mu\nu} \mathcal{L}_m.
\end{equation}
In order to calculate the variation of the energy-momentum tensor with respect to the metric, a new tensor is defined, i.e.
\begin{equation}
    \Theta_{\mu\nu} \equiv \frac{\delta T_{\alpha\beta}}{\delta g^{\mu\nu}} g^{\alpha\beta}.
\end{equation}
Using eq.(\ref{tensor energia}), this tensor becomes
\begin{equation}
    \Theta_{\mu\nu}= -2T_{\mu\nu} + g_{\mu\nu} \mathcal{L}_m  -2g^{\alpha\beta} \frac{\partial^2 \mathcal{L}_m}{\partial g^{\mu\nu}\partial g^{\alpha\beta}}.
\end{equation}

Varying the action (\ref{action_1}) with respect to the metric $g^{\mu\nu}$ and assuming that $\delta \tilde{R}_{\mu\nu}\pc{\tilde{\Gamma}}=0$, the field equations are given as
\bea\label{field_1}
     \tilde{R}_{\mu\nu}f_R = \kappa^2 T_{\mu\nu} - \pc{T_{\mu\nu} + \Theta_{\mu\nu}} f_T + \frac{g_{\mu\nu}}{2} f, 
\eea
where $f_R \equiv \frac{\partial f}{\partial R}$ and $f_T \equiv \frac{\partial f}{\partial T}$. 

An interesting constraint, that simplifies the field equations, comes from the trace of the equation. The trace of eq. (\ref{field_1}) is
\begin{equation}\label{field_2}
    R\left(g,\tilde{\Gamma}\right) f_R = \kappa^2 T - \pc {T + \Theta} f_T + 2f,
\end{equation}
with $\Theta=\Theta^\mu\,_\mu$. Combining eq. (\ref{field_1}) and eq. (\ref{field_2}), field equations are written as
\begin{equation}\label{conteudo_materia}
     f_RG_{\mu\nu} \pc{g, \tilde{\Gamma}} =  \kappa^2 T_{\mu\nu} - f_T\pc{T_{\mu\nu} + \Theta_{\mu\nu}} - \frac{1}{2} \pr{f+\kappa^2 T - f_T\pc{ T+\Theta}}g_{\mu\nu},
\end{equation}
where $G_{\mu\nu} \pc{g, \tilde{\Gamma}}$ is the Einstein tensor in the Palatini formalism, which is defined as
\bea
    G_{\mu\nu} \pc{g, \tilde{\Gamma}} = \tilde{R}_{\mu\nu} \pc{\tilde{\Gamma}} - \frac{1}{2} g_{\mu\nu} \tilde{R} \pc{g,\tilde{\Gamma}}.
\eea

As the action depends on the metric and the Palatini connection, now varying it with respect to the connection $\tilde{\Gamma}$, keeping the metric constant, leads to
\begin{equation}
    \delta S = \frac{1}{\kappa ^2} \int \sqrt{-g} f_R g^{\mu\nu} \pr{\tilde{\nabla} \pc{\delta \tilde{\Gamma}_{\mu\nu}^{\lambda}} - \tilde{\nabla} \pc{\delta \tilde{\Gamma}_{\mu\lambda}^{\lambda}}} d^4x.
\end{equation}
where was it used that
\begin{align}
    \delta f\left(R(\tilde{\Gamma})\right)  &= f_R g^{\mu\nu} \pr{\tilde{\nabla} \pc{\delta \tilde{\Gamma}_{\mu\nu}^{\lambda}} - \tilde{\nabla} \pc{\delta \tilde{\Gamma}_{\mu\lambda}^{\lambda}}},
\end{align}
with $\tilde{\nabla}$ being the covariant derivative associated with $\tilde{\Gamma}$. 

Defining $A^{\mu\nu}\equiv f_Rg^{\mu\nu}$ and integrating by parts, the action variation becomes
\begin{equation}
    \kappa^2 \delta S = \int \tilde{\nabla}_\lambda \pr{\sqrt{-g} \pc{A^{\mu\nu} \delta \tilde{\Gamma}_{\mu\nu}^{\lambda} -A^{\mu\lambda} \delta \tilde{\Gamma}_{\mu\alpha}^{\alpha}}}d^4x - 
    \int \tilde{\nabla}_\lambda \pr{\sqrt{-g} \pc{A^{\mu\nu} \delta^\lambda_\alpha - A^{\mu\lambda} \delta^\nu_\alpha}} \delta \tilde{\Gamma}_{\mu\nu}^{\alpha} d^4x.\label{Gauss}
\end{equation}
Note that, the first term in eq. (\ref{Gauss}) is a total derivative. Using Gauss theorem, it is possible to transform it into a surface integral and then it vanishes. Thus, the variation of the action yields
\begin{equation}\label{integral}
    \tilde{\nabla}_\lambda \pr{ \sqrt{-g} \pc { A^{\mu\nu} \delta^\lambda_\alpha - A^{\mu\lambda} \delta^\nu_\alpha}} = 0.  
\end{equation}
Considering the case $\lambda \neq \alpha$, eq.(\ref{integral}) is written as
\begin{equation}\label{conformal_1}
    \tilde{\nabla}_\lambda \pr{ \sqrt{-g} f_R g^{\mu\nu}} =0.
\end{equation}
The eq. (\ref{conformal_1}) shows that the connection $\tilde{\Gamma}$ is compatible with a conformal metric $\tilde{g}_{\mu\nu}=f_R g_{\mu\nu}$. This implies that the geometry is not modified and the Palatini connection may be written as
\begin{equation}
    \tilde{\Gamma}_{\mu\nu}^{\lambda} = \frac{1}{2} \tilde{g}^{\lambda\rho} \pc{\partial_\nu\tilde{g}_{\rho\mu} + \partial_\mu\tilde{g}_{\rho\nu} + \partial_\rho\tilde{g}_{\mu\nu}}.
\end{equation}

The connection $\tilde{\Gamma}$ can be written in terms of the Levi-Civita connection $\Gamma$, since the metrics $g_{\mu\nu}$ and $\tilde{g}_{\mu\nu}$ are conformally related. Then
\begin{equation}
    \tilde{\Gamma}_{\mu\nu}^{\lambda} = \Gamma_{\mu\nu}^{\lambda} + \frac{1}{2f_R} \pc{ \delta^\lambda_\mu \partial_\nu + \delta^\lambda_\nu \partial_\mu - g_{\mu\nu} \partial^\lambda } f_R.
\end{equation}

Using this relation and the properties of a conformal metric, the Ricci tensor is given as
\begin{equation}
    \tilde{R}_{\mu\nu}=R_{\mu\nu} + \frac{1}{f_R} \pr{\frac{3}{2F_r} \nabla_\mu f_R \nabla_\nu f_R - \pc{\nabla_\mu \nabla_\nu + \frac{g_{\mu\nu}}{2} \Box } f_R}.
\end{equation}
Note that, the tilde quantities are associated with the conformal metric. In a similar way the Ricci scalar and the Einstein tensor in the conformally related frames are given, respectively, as
\begin{equation}
    \tilde{R} = R - \frac{3}{f_R} \Box f_R + \frac{3}{2f_R} \pc{\nabla f_R},
\end{equation}
and
\begin{equation}
    \tilde{G}_{\mu\nu} = G_{\mu\nu} + \frac{1}{f_R} \chav { \pc{g_{\mu\nu} \Box - \nabla_\mu \nabla_\nu} f_R + \frac{3}{2f_R} \pr{\nabla_\mu f_R \nabla_\nu f_R - \frac{g_{\mu\nu}}{2} \pc{\nabla f_R}^2 }}.   
\end{equation}

Therefore, the field equations, eq. (\ref{conteudo_materia}), for Palatini $f(R,T)$ gravity theory becomes
\begin{align}\label{eq_campo1}
    G_{\mu\nu} + \frac{1}{f_R} \chav { \pc{g_{\mu\nu} \Box - \nabla_\mu \nabla_\nu} f_R + \frac{3}{2f_R} \pr{\nabla_\mu f_R \nabla_\nu f_R - \frac{g_{\mu\nu}}{2} \pc{\nabla f_R}^2 }} = \\ \nonumber \ \frac{1}{f_R} \chav {8\pi \pc{T_{\mu\nu} -\frac{g_{\mu\nu}}{2} T} + \pr{\pc{T_{\mu\nu} + \Theta_{\mu\nu} } -\frac{g_{\mu\nu}}{2} \pc{T + \Theta} }f_T - \frac{g_{\mu\nu}}{2} }.
\end{align}
It is important to note that, eq. (\ref{eq_campo1}) is expressed only in the $g$ frame, i.e. it is written in terms of the metric $g_{\mu\nu}$, its derivatives, and the matter fields.

In the next section, this field equation is considered and the causality violation for different matter contents is studied.

\section{G\"{o}del Metric in Palatini $f(R,T)$ gravity theory}

To investigate a possible violation of causality in Palatini $f(R,T)$ gravity theory, the G\"{o}del metric is considered. This solution was proposed by Kurt G\"{o}del, in 1949 \cite{godel}. It is an exact solution of Einstein equations, for a homogeneous rotating universe. As a consequence the possibility of Closed Timelike Curves (CTCs) emerge. CTCs allow the violation of causality and, theoretically, permit time-travel in this space-time. Its line element is defined as
\begin{equation}\label{godel_metric}
    ds^2 = a^2 \left(dt^2 - dx^2 + \frac{e^{2x}}{2}dy^2 - dz^2 + 2e^xdt \ dy \right),
\end{equation}
where $a$ is an arbitrary number. 

In order to study the field equations of Palatini $f(R,T)$ gravity, some tensor quantities associated with G\"{o}del space-time are calculated. The non-zero Ricci tensor components are
\begin{align}
    R_{00} = 1, \ \ \ R_{02} = R_{20} = e^x, \ \ \ R_{22} = e^{2x},
\end{align}
and the Ricci scalar is
\begin{equation}
    R = \frac{1}{a^2}.
\end{equation}

Another important ingredient in this study is the matter content. Let us take the perfect fluid as a matter content. Its energy-momentum tensor is
\begin{equation}
    T_{\mu\nu} = \pc {\rho + p}u_\mu u_\nu -pg_{\mu\nu},
\end{equation}
where $u_\mu$ is the quadri-velocity tensor whose covariant components are $u_\mu= (a,0,ae^x,0)$, $\rho$ and $p$ are energy density and pressure, respectively. Here, the case $p=0$ is considered. In fact, the energy-momentum tensor describes only dust as matter content. Then
\begin{align}
    T_{\mu\nu} &= \rho u_\mu u_\nu,
\end{align}
and the trace is
\begin{align}
    T &= \rho.
\end{align}
In the same way, the tensor $\Theta_{\mu\nu}$ for a pressureless perfect fluid becomes
\begin{equation}
    \Theta_{\mu\nu} = -2T_{\mu\nu}.\label{Theta}
\end{equation}

Then field equations, eq. (\ref{eq_campo1}), are written as
\begin{equation}\label{eq_campo_2}
    G_{\mu\nu} + J_{\mu\nu} =  \ \frac{1}{f_R} \chav {8\pi \pc{T_{\mu\nu} -\frac{g_{\mu\nu}}{2} T} + \pr{\pc{T_{\mu\nu} + \Theta_{\mu\nu} } -\frac{g_{\mu\nu}}{2} \pc{T + \Theta} }f_T - \frac{g_{\mu\nu}}{2} },
\end{equation}
where
\begin{equation} \label{tensor_J}
   J_{\mu\nu} =  \frac{1}{f_R} \chav { \pc{g_{\mu\nu} \Box - \nabla_\mu \nabla_\nu} f_R + \frac{3}{2f_R} \pr{\nabla_\mu f_R \nabla_\nu f_R - \frac{g_{\mu\nu}}{2} \pc{\nabla f_R}^2 }}.
\end{equation}
Note that the Ricci scalar for the G\"{o}del metric takes a constant value. This implies $J_{\mu\nu} = 0$. Then the field equations for this metric are given as
\begin{align}
    \frac{1}{f_R} \chav{4\pi \rho a^2 + \frac{3}{2} \rho a^2 f_T -\frac{a^2}{f} } + \frac{1}{2} -a^2\Lambda & = 0,\\
    \frac{1}{f_R} \chav{4\pi \rho a^2 + \frac{1}{2} \rho a^2 f_T -\frac{a^2}{f} } - \frac{1}{2} -a^2\Lambda & = 0,\\ \label{resolvendo1}
    \frac{1}{f_R} \chav{4\pi \rho a^2 + \frac{1}{2} \rho a^2 f_T +\frac{a^2}{f} } - \frac{1}{2} -a^2\Lambda & = 0,\\ \label{resolvendo2}
    \frac{1}{f_R} \chav{24 \pi \rho a^2 + 3 \rho a^2 f_t - a^2f} - 2 + a^2\Lambda &=0.
\end{align}
From eq. (\ref{resolvendo1}) and eq. (\ref{resolvendo2}) we get
\begin{equation}
    \rho = \frac{f_R}{a^2 \pc {8\pi + f_T}}
\end{equation}
and 
\begin{equation}
    \Lambda = -\frac{f}{2f_R}.
\end{equation}
These results show that G\"{o}del metric is a solution of Palatini $f(R,T)$ gravity theory. It means that this theory allows causality violation. So, the possibility of CTCs is real and time travel to the past is theoretically possible. In addition, this is a generalization of GR solution that is recovered in the limit $f = R$ ($f_R \rightarrow 1$) and $f_T = 0$.  In the next section, this phenomenon is investigated in more detail.

\section{G\"{o}del-type Metric in Palatini $f(R,T)$ gravity theory}

A generalization of the G\"{o}del solution, called G\"{o}del-type solution, has been developed \cite{tipo_godel1}. Its line element is 
\begin{equation}\label{type_godel}
    ds^2 = -dt^2 - 2H(r) dtd\phi + dr^2 + G(r) d\phi^2 + dz^2, 
\end{equation}
where $H(r)$ and $D(r)$ are functions defined as \cite{frt_and_godel} 
\begin{align}
    H(r) &= \frac{4\omega}{m^2} \sinh^2 \left(\frac{mr}{2} \right),\label{condition1} \\
     D(r) &= \frac{1}{m} \sinh(mr)\label{condition2},
\end{align}
and $G(r) = D^2(r) - H^2(r)$. The parameters $\omega$ and $m$ are such that $\omega^2 > 0$ and $-\infty \le m^2  \le +\infty$. The G\"{o}del-type metrics are defined by the two parameters $\omega$ and $m$. Note that, identical pairs $(\omega, m)$ specify isometric space-times \cite{tipo_godel1}. The standard G\"{o}del solution is a particular case of the $m^2>0$ class of space-times in which $m^2=2\omega^2$. However, causal and non-causal regions are allowed. These regions are determined from free parameters of the metric $m$ and $\omega$, and limited by a critical radius $r_c$. From this critical radius, there is a violation of causality. It is defined as
\begin{equation}
    \sinh^2 \left(\frac{mr_c}{2}\right) \ = \ \left(\frac{4\omega^2}{m^2} -1 \right)^{-1}.
\end{equation}
It is important to note that, for $m^2=4\omega^2$ the critical radius becomes infinite $(r_c=\infty)$, this implies a causal universe. For $m^2 \ge 4\omega^2$, there are no G\"{o}del-type CTCs, and the breakdown of causality is avoided. Therefore, the G\"{o}del-type solution brings more details to the problem of causality. 

In order to solve the field equations, a new basis, for simplicity, is chosen \cite{tipo_godel1}. The metric becomes
\begin{equation}
    ds^2 = \eta_{AB} \theta^A \theta^B = (\theta^0)^2 - (\theta^1)^2 - (\theta^2)^2 - (\theta^3)^2, \label{frame}
\end{equation}
where Latin letters denote the transformed space. The one-forms $\theta^A$ is defined as
\begin{equation}
    \theta^A = e^A\ _{\mu}\ dx^\mu,  
\end{equation}
and its components are
\begin{align}
    \theta^{(0)} &= dt + H(r)d\phi, \\ \theta^{(1)} &= dr, \\ \theta^{(2)} &= D(r)d\phi, \\ \theta^{(3)} &= dz,    
\end{align}
with $e^A\ _{\mu}$ being the tetrads, such that $ e^A\,_\mu e^\mu\,_B=\delta^A_B$. The non-null components of the tetrads are
\begin{equation}
        e^{0}\ _{(0)} = e^{1}\ _{(1)} = e^{3}\ _{(3)} = 1, \quad e^{0}\ _{(2)} = - \frac{H(r)}{D(r)}, \quad e^{2}\ _{(2)} = D^{-1}(r).
\end{equation}

Using that $G_{AB}=e^\mu_A e^\nu_B G_{\mu\nu}$, the non-vanishing components of the Einstein tensor in the flat (local) space-time take the form
\begin{align}
     G_{(0)(0)} &= 3\omega^2-m^2, \\
    G_{(1)(1)} &= \omega^2,\\
    G_{(2)(2)} &= \omega^2, \\
    G_{(3)(3)} &= m^2-\omega^2.
\end{align}

Assuming that the content of matter is a perfect fluid, whose its energy-momentum tensor is given by
\begin{equation}
    T_{AB}=(\rho+p)u_A u_B-p\eta_{AB},
\end{equation}
where $u_A = (1,0,0,0)$, the field equations for Palatini $f(R,T)$ gravity theory, in the tangent (flat) space, i.e.
\begin{equation}\label{eq_campo_tipo_godel}
    G_{AB} + J_{AB} =  \ \frac{1}{f_R} \chav {8\pi \pc{T_{AB} -\frac{\eta_{AB}}{2} T} + \pr{\pc{T_{AB} + \Theta_{AB} } -\frac{\eta_{AB}}{2} \pc{T + \Theta} }f_T - \frac{\eta_{AB}}{2} },
\end{equation}
provides the set of equations
\begin{align}
    2f_R\pc{3\omega^2 - m^2} +f  &= 8\pi \pc {\rho + 3p} + \pc{\rho +p}f_T, \\
    2f_R\omega^2 - f &= 8\pi \pc {\rho - p} + \pc{\rho +p}f_T, \label{resolvendo_tipo1}\\ 
    2f_R\pc{m^2 - \omega^2} -f &= 8\pi \pc {\rho - p} + \pc{\rho +p}f_T. \label{resolvendo_tipo2}
\end{align}
Here $J_{AB}=0$, since the Ricci scalar for the G\"{o}del-type metrics assumes a constant value, i.e. $R=2(m^2-\omega^2)$. From eq. (\ref{resolvendo_tipo1}) and eq. (\ref{resolvendo_tipo2}) we get,
\begin{equation}
    m^2=2\omega^2.\label{GC}
\end{equation}
This condition defines the original G\"{o}del universe. Using this result, the remaining equations become
\begin{align}
    f_R m^2 + f &= 8\pi \pc{\rho +3p } + \pc{\rho +p} f_T,\\
     f_R m^2 - f &=8\pi \pc{\rho -p } + \pc{\rho +p} f_T.
\end{align}
Taking these equations, the $m$ parameter is written as
\begin{equation}
    m^2 = \frac{1}{16\pi f_R}\left[(8\pi+f_T)(16\pi\rho+f)\right]. 
\end{equation}
Then, the critical radius (beyond which the causality is violated) becomes 
\begin{equation}\label{raio_critico1}
    r_c = 2\sinh^{-1}(1)\sqrt{\frac{16\pi f_R}{(8\pi+f_T)(16\pi\rho+f)}}.
\end{equation}
This result leads to the following understanding: (i) the G\"{o}del-type metric is a solution of the Palatini $f(R,T)$ gravity theory; (ii) for the case $f_T = 0$, the eq. (\ref{raio_critico1}) yields the result obtained for $f(R)$ gravity \cite{tipo_godel_fR} and for $f(R,T)\rightarrow f(R)=R$, the standard GR results are recovered; (iii) the critical radius depends on the gravity theory (that is, on the function $f$ and its derivatives), and density of matter $\rho$. Note that, as the G\"{o}del condition is present, eq. (\ref{GC}), there is a non-causal region beyond the critical radius. This condition implies that the result for a perfect fluid as the content of matter is necessarily isometric to the G\"{o}del geometry, i.e. unavoidably exhibit closed timelike curves. Now, a question arises: what are the conditions for obtaining a causal universe in this gravitational theory? In order to obtain such conditions, different types of matter content are investigated.

\subsection{Perfect fluid with scalar field}

In this subsection, in order to obtain more information about causality in the Palatini $f(R,T)$ gravity theory, the scalar field and the perfect fluid are considered as matter content. The total energy-momentum tensor is composed by two parts, the scalar field $T_{AB}^S$ and the perfect fluid $T_{AB}^M$. Then
\begin{align}
    T_{AB} &= T_{AB}^M + T_{AB}^S, \\ 
    &= \pc{\rho + p} u_Au_B -p\eta_{AB} + \nabla_A \phi \nabla_B \phi - \frac{1}{2} \eta_{AB}\eta^{CD}\nabla_C \phi \nabla_D \phi,
\end{align}
where $\nabla_A$ is a covariant derivative that has a basis as $\theta^A = e^A_\beta dx^\beta$. The scalar field is given as $\phi = \epsilon z + \epsilon $, with $\epsilon = const$. This condition satisfies the scalar field equation $\Box \phi = \eta^{AB} \nabla_A\nabla_B \phi = 0$. The non-zero components of the energy-momentum tensor associated with the scalar field are
\begin{equation}
    T_{00}^S = -T_{11}^S = -T_{22}^S = T_{33}^S = \frac{\epsilon^2}{2}.
\end{equation}
The trace of the energy-momentum tensor is 
\begin{equation}
    T=T^M+T^S=\rho -3p+\epsilon^2.
\end{equation}
Considering the scalar field contributions, the tensor $\Theta_{AB}$ is written as
\begin{equation}
    \Theta_{AB} = \Theta_{AB}^M + \Theta_{AB}^S,
\end{equation}
where $\Theta_{AB}^M$ is given in eq. (\ref{Theta}). By taking the free Lagrangian from the scalar field
\begin{equation}
    \mathcal{L}^S = \eta^{AB} \nabla_A \phi \nabla_B \phi, 
\end{equation}
the tensor $\Theta_{AB}^S$ is obtained as
\begin{equation}
    \Theta_{AB}^S = - T_{AB}^S + \frac{1}{2} T^S \eta_{AB}.
\end{equation}
Then the field equation eq.(\ref{eq_campo_tipo_godel}) becomes
\begin{align}
    f_R G_{AB} &= 8\pi \pr {\pc{\rho  + p}u_Au_B -p\eta_{AB} + T_{AB}^S }\\
    &-\frac{1}{2} \pr{8\pi \pc{\rho - 3p + \epsilon^2 } +f +f_T \pc{\rho + p - 2\eta^2 } }\eta_{AB}\\
    & +f_T \pr{\pc{\rho + p }u_Au_B - \frac{1}{2} \epsilon^2 \eta_{AB}}.
\end{align}
This equation leads to the set of equations
\begin{align}
    8\pi \epsilon^2 &= \pc{m^2 - 2\omega^2} f_R,\\
    8\pi + \frac{1}{2}\epsilon^2 f_T &= \frac{1}{2} \pc{2\omega^2 - m^2}f_R +\frac{1}{2} f,\\
    8\pi \rho + f_T \pc{ \rho + p -\frac{1}{2}\epsilon^2} &= \frac{1}{2} \pc{6\omega^2 - m^2} f_R -\frac{1}{2}f,
\end{align}
where $f_R=\partial f / \partial R > 0$ and $f_T = \partial f / \partial T > 0$ have been considered. These equations allow a causal solution that is given as
\bea
m^2&=&4\omega^2\\
f_R &=& \frac{8\pi \epsilon^2}{2\omega^2}.
\eea
This implies a critical radius $r_c \rightarrow \infty$. As a consequence, for this combination of matter fields, the violation of causality is not permitted. In other words, a completely causal universe in this gravitational model is possible and depends on the matter content.

\subsection{Scalar field}

Here only the scalar field $\phi(z) = \epsilon z + \epsilon $ is considered to be matter content. In this case, the field equations are
\begin{align}
    f_R\pc{3\omega^2 - m^2} + \frac{f}{2} &= \frac{1}{2}\epsilon^2 f_T,\\
    f_R\omega^2 - \frac{f}{2} &= - \frac{1}{2} \epsilon^2 f_T,\\
    f_R\pc{m^2 - \omega^2} -\frac{f}{2} &= 8\pi \epsilon^2 -\frac{1}{2} f_T \epsilon^2. 
\end{align}
Considering the conditions $f_R > 0$ and $f_T > 0$ the causal condition $m^2= 4\omega^2$ is obtained for any function $f(R,T)$. Then, a causal universe with a single scalar field as matter content is allowed. Therefore, these results show the causality problem in this formalism depends on two main ingredients: the gravitational theory and the matter source.

In order to emphasize, the study developed here, it is important to note that, cosmological solutions in $f(R,T)$ gravity theory have been investigated in both formalisms, i.e. metric and Palatini. In the Palatini formulation, the field equations contain extra terms generated by the coupling between the trace of the energy-momentum tensor and geometry. In this formalism, the energy-momentum tensor of the matter is not conserved and the motion of the particles is not geodesic as in the metric case. In addition, due to the matter-geometry coupling an extra force arises. Then,  no new physics is expected in the motion of massive test particles in the Palatini formulation of the $f(R, T)$ gravity. However, for the Friedmann-Robertson-Walker universe the cosmological equations in the Palatini formalism are different due to the presence of some dynamical terms associated to $f_R$ \cite{fRT_Palatini_primeiro,palatini_fRT1}. Therefore, in this case, the Palatini $f(R, T)$ theory allows a much richer cosmological dynamics, as compared to the metric formulation. In this paper other cosmological models are examined. The G\"{o}del and G\"{o}del-type models are investigated and the problem of causality violation in Palatini $f(R, T)$ gravity generalizes the results obtained in Refs. \cite{godel_fRT, frt_and_godel}.

\section{Conclusion}

In the present paper the $f(R,T)$ gravity theory is considered. It was shown that this theory generalizes GR, implying a geometry-matter coupling, with the trace of the energy-momentum tensor included as a field variable in the gravitational action. Here, the Palatini formalism is introduced and then $f(R,T)$ gravity theory is formulated in this approach. In the Palatini formulation, the metric and affine connection are taken as independent field variables. Based on observational data, there are some motivations for studying modified theories of gravity like $f(R,T)$, because it provides an alternative way to explain the acceleration of the universe or leads to a possible quantum description of gravity. If $f(R,T)$ is a model that describes gravity, a number of questions should be reexamined in this framework. In this paper, the causality problem is investigated in Palatini $f(R,T)$ gravity theory. Our results show that the G\"{o}del metric is a solution in this gravitational model. Then the violation of causality is permitted. Furthermore, to generalize the problem of causality, G\"{o}del-type metric with perfect fluid as matter content is explored. In this case, the usual condition for G\"{o}del universe is obtained. In addition, a critical radius is calculated. The result has shown that the modification in the critical radius depends on the gravitational theory $f$, its derivatives and matter content. In order to find a causal solution, a combination of perfect fluid and scalar field as matter content has been considered. In this case, there is a solution for which the critical radius becomes infinity. Therefore, a causal solution is allowed. By considering a single scalar field as a matter source, similar causal solution has been obtained. Then for a well-motivated matter source, causal and non-causal solutions in Palitini $f(R,T)$ gravity theory are allowed. Furthermore, it is important to note that, if gravity is governed by Palatini $f(R,T)$ theory various issues of both observational and theoretical nature ought to be reexamined in this framework, including the question as to whether these theories permit the breakdown
of causality at a non-local scale. Therefore, our result is an important theoretical test for this gravitational theory. Since it is a generalization of GR, it must contain all the standard and exact solutions of GR.

\section*{Acknowledgments}

This work by A. F. S. is supported by CNPq projects 308611/2017-9 and 430194/2018-8; J. S. Gon\c{c}alves thanks CAPES for financial support.


\begin{thebibliography}{99}

\bibitem{expansao_ace} G. Goldhaber and S. Perlmutter, Phys. Rep. {\bf 307},  325 (1998).
\bibitem{expansao_ace2} S. Perlmutter et al., Astrophys. J. {\bf 517}, 565 (1999).
\bibitem{WMAP2007} D. Spergel et al., Astrophys. J. Suppl {\bf 170}, 377 (2007).
\bibitem{WMAP2009} E. Komatsu et al., Astrophys. J. Suppl. {\bf 180}, 330 (2009).
\bibitem{WMAP2011} E. Komatsu, K. M. Smith, J. Dunkley, C. L. Bennett, B. Gold, G. Hinshaw, N. Jarosik, D. Larson, M. R. Nolta, L. Page, and et al., Astrophys. J. Suppl. {\bf 192} (2011).
\bibitem{LSS_evidence} G. E. Addison, G. Hinshaw, and M. Halpern, Mon. Not. R. Astron. Soc. {\bf 436}, 1674 (2013).
\bibitem{weak_lensing} B. Jain and A. Taylor, Phys. Rev. Lett. {\bf 91}, 141302 (2003).
\bibitem{dark_energy1} R. R. Caldwell and M. Kamionkowski, Annual Review of Nuclear and Particle Science {\bf 59}, 397 (2009).
\bibitem{dark_energy2} M. Li, X.-D. Li, S. Wang, and Y. Wang, Communications in Theoretical Physics {\bf 56}, 525 (2011).
\bibitem{k_essece_1} T. Chiba, T. Okabe, and M. Yamaguchi, Phys. Rev. D {\bf 62}, 023511 (2000).
\bibitem{review1} T. Clifton, P. G. Ferreira, A. Padilla and C. Skordis, 	Phys. Rep. {\bf 513}, 1 (2012).
\bibitem{review2} R. Myrzakulov, L. Sebastiani and S. Zerbini, Int. J. Mod. Phys. D {\bf 22}, 1330017 (2013).
\bibitem{review3} A. Felice and S. Tsujikawa, Living Rev. Relativity {\bf 13}, 3 (2010).
\bibitem{review4} T. P. Sotiriou and V. Faraoni, Rev. Mod. Phys. {\bf 82}, 451 (2010).
\bibitem{RL}T. Harko and F. S. N. Lobo, Eur. Phys. J. C {\bf 70}, 373 (2010).
\bibitem{RL1} T. Harko and F. S. N. Lobo, Galaxies {\bf 2}, 410 (2014).
\bibitem{Harko} T. Harko, F. S. N. Lobo, S. Nojiri and S. D. Odintsov, Phys. Rev. D {\bf 84}, 024020 (2011).
\bibitem{QT} R.-J. Yang, Physics of the Dark Universe {\bf 13}, 87 (2016).
\bibitem{fRT1} S. I. Vacaru, E. V. Veliev, and E. Yazici, Int. J. Geom. Meth. Mod. Phys. {\bf 11}, 1450088 (2014).
\bibitem{fRT2} M. Farasat Shamir and Z. Raza, Astrophys. Space Sci. {\bf 356}, 111 (2015).
\bibitem{fRT3} C. P. Singh and V. Singh, {\it Friedmann cosmology with particle creation in modified f(R; T) gravity}, arXiv:1408.0633 [gr-qc].
\bibitem{Houndjo} M. S. J. Houndjo, Int. J. Mod. Phys. D {\bf 21}, 1250003 (2012).
\bibitem{Momeni} D. Momeni, R. Jamil and R. Myrzakulov, Eur. Phys. J. C {\bf 72}, 1999 (2012).
\bibitem{Shabani} H. Shabani and M. Farhoud, Phys. Rev. D {\bf 90}, 044031 (2014), arXiv:1407.6187. 
\bibitem{Sharif} M. Sharif and M. Zubair, J. High Energy Phys. {\bf 12}, 079 (2013).
\bibitem{Kiani} F. Kiani and K. Nozari, K., Phys. Lett. B {\bf 728}, 554 (2014).
\bibitem{Sharif2} M. Sharif and M. Zubair, J. Cosmol. Astropart. Phys. {\bf 03}, 028 (2012).
\bibitem{Jamil} M. Jamil, D. Momeni and M. Ratbay, Chin. Phys. Lett. {\bf 29}, 109801 (2012).
\bibitem{Sharif3} M. Sharif and M. Zubair, J. Exp. Theor. Phys. {\bf 117}, 248 (2013).
\bibitem{Azizi} T. Azizi, Int. J. Theor. Phys. {\bf 52}, 3486 (2013).
\bibitem{Alvaro} F. G. Alvarenga, A. de la Cruz-Dombriz, M. J. S. Houndjo, M. E. Rodrigues and D. S\'{a}ez-G\'{o}mez, Phys. Rev. D {\bf 87}, 103526 (2013), arXiv:1302.1866. 
\bibitem{Reddy} D. R. K. Reddy, R. L. Naidu, K. Dasu Naidu and T. Ram Prasad, Astrophys. Space Sci. {\bf 346}, 261 (2013).
\bibitem{Moraes} P. H. R. S. Moraes, Astrophys. Space Sci. {\bf 352} 273 (2014).
\bibitem{Einstein1} A. Einstein, Sitzungsberichte der K\"{o}niglich Preussischen Akademie der Wissenschaften (Berlin) {\bf 1923}, 76 (1923).
\bibitem{Einstein2} A. Einstein, Sitzungsberichte der K\"{o}niglich Preussischen Akademie der Wissenschaften (Berlin) {\bf 137}, 32 (1923).
\bibitem{palatini_motivation1} G. J. Olmo, Int. J. Mod. Phys. D {\bf 20}, 413 (2011).
\bibitem{fRT_Palatini_primeiro} E. Barrientos, F. S. Lobo, S. Mendoza, G. J. Olmo, and D. Rubiera-Garcia, Phys. Rev. D {\bf 97}, 104041 (2018).
\bibitem{palatini_fRT1} J. Wu, G. Li, T. Harko, and S.-D. Liang, Eur. Phys. J. C {\bf 78}, 430 (2018).
\bibitem{godel} K. G\"{o}del, Rev. Mod. Phys. {\bf 21}, 447 (1949).
\bibitem{ctc1} J. R. Gott, Phys. Rev. Lett. {\bf 66}, 1126 (1991).
\bibitem{ctc2} R. P. Kerr, Phys. Rev. Lett. {\bf 11}, 237 (1963).
\bibitem{tipo_godel1} M. J. Rebouças and J. Tiomno, Phys. Rev. D {\bf 28}, 1251 (1983).

\bibitem{fr_and_godel} M. J. Rebouças and J. Santos, Phys. Rev. D {\bf 80}, 063009 (2009).
\bibitem{kessence_and_godel} J. G. da Silva and A. F. Santos,  Eur. Phys. J. Plus {\bf 135}, 1 (2020).
\bibitem{chersimon_and_godel1} C. Furtado, T. Mariz, J. R. Nascimento, A. Y. Petrov, and A. F. Santos, Phys. Rev. D {\bf 79}, 124039 (2009).
\bibitem{chersimon_and_godel2} C. Furtado, J. Nascimento, A. Petrov, and A. F. Santos, Phys. Lett. B {\bf 693}, 494 (2010).
\bibitem{ft_and_godel} G. Otalora and M. J. Rebouças, Eur. Phys. J. C {\bf 77}, 799 (2017).
\bibitem{frt_and_godel} A. F. Santos and C. Ferst, Mod. Phys. Lett. A {\bf 30},  1550214 (2015).
\bibitem{bumblebeee_and_godel} J. R. Nascimento, A. Y. Petrov, A. F. Santos, and W. D. R. Jesus, Mod. Phys. Lett. A {\bf 30}, 1550011 (2015).
\bibitem{horava_and_godel} J. Fonseca-Neto, A. Petrov, and M. Rebouças, Phys. Lett. B {\bf 725}, 412 (2013).
\bibitem{brans_and_godel} J. Agudelo, J. Nascimento, A. Petrov, P. Porfírio, and A. F. Santos, Phys. Lett. B {\bf 762}, 96 (2016).
\bibitem{frq_and_godel} F. Gama, J. Nascimento, A. Petrov, P. Porfírio, and A. F. Santos, Phys. Rev. D {\bf 96}, 064020 (2017).
\bibitem{godel_fRT} A. F. Santos, Mod. Phys. Lett. A {\bf 28}, 1350141 (2013).
%\bibitem{tipo_godel_frt} A. F. Santos and C. J. Ferst, Mod. Phys. Lett. A {\bf 30}, 1550214 (2015).
\bibitem{tipo_godel_fR} J. Santos, M. J. Reboucas, and T. B. R. F. Oliveira, Phys. Rev., vol. D {\bf 81}, 123017 (2010).
\bibitem{palatini_fR} J. Santos, M. J. Rebouças and A. F. F. Teixeira, Eur. Phys. J. C {\bf 78}, 567 (2018).

\end{thebibliography}
\end{document}